\documentclass[letterpaper,10pt,twocolumn,twoside]{IEEEtran}

\usepackage{graphicx}
\usepackage{amsmath}
\usepackage{amssymb}
\usepackage{setspace}
\usepackage{verbatim}
\usepackage{algorithmic}
\usepackage{bbm}
\usepackage{multirow}
\usepackage{psfrag}
\usepackage{booktabs}
\usepackage{cite}

\newtheorem{thm}{Theorem}
\newtheorem{cor}[thm]{Corollary}

\newtheorem{defn}[thm]{Definition}
\newtheorem{ex}[thm]{Example}
\newtheorem{alg}[thm]{Algorithm}
\newtheorem{asm}[thm]{Assumption}

\newcommand{\bX}{\mathbf{X}}

\newcommand{\bR}{\mathbf{R}}
\newcommand{\bp}{\mathbf{p}}

\newcommand{\bw}{\mathbf{w}}
\newcommand{\btau}{\mathbf{\tau}}

\newcommand{\Xh}{\widehat{X}}
\newcommand{\Sh}{\widehat{S}}

\newcommand{\eps}{\epsilon}

\newcommand{\cN}{\mathcal{N}}

\newcommand{\sni}{S_{t_i}}
\newcommand{\htsni}{\hat{\theta}_{\sni}}

\newcommand{\maxervar}{\sigma_{\rm max}^2}

\newcommand{\rh}{\hat{r}}

\def\argmin{\mathop{\mathrm{arg\,min}}}

\def\arginf{\mathop{\mathrm{arg\,inf}}}
\def\cmin{c_{\rm min}}
\def\ctot{c_{\rm tot}}
\def\dtot{d_{\rm tot}}
\def\Ttot{T_{\rm tot}}
\def\Troot{T_{\rm root}}
\def\Tsgreedy{T_{\rm greedy}^*}

\newcommand{\Frac}[2]{{{#1}/{#2}}}

\begin{document}
\title{Time-Stampless Adaptive Nonuniform Sampling for Stochastic Signals}

\author{
Soheil Feizi, Vivek K Goyal, \emph{Senior Member, IEEE}, and
  Muriel M\'edard, \emph{Fellow, IEEE}%
  \thanks{This material is based upon work supported in part by the
    National Science Foundation under Grant No.\ 0643836, the Texas Instruments Leadership University Program and AFOSR under award No. 016974-002.
    The material in this paper was present in part at 2010 Annual Allerton Conference on Communication, Control, and Computing.}
  \thanks{The authors are with the Research Laboratory of Electronics and
    the Department of Electrical Engineering and Computer Science at the
    Massachusetts Institute of Technology, Cambridge, MA 02139 USA
    (email: \{sfeizi,vgoyal,medard\}@mit.edu).
    Telephone: +1~617~253~3167.}
}

\maketitle

\begin{abstract}
In this paper, we introduce a \emph{time-stampless adaptive nonuniform sampling} (TANS) framework, in which time increments between samples are determined by a function of the $m$ most recent increments and sample values. 
Since only past samples are used in computing time increments, it is not necessary to save sampling times (time stamps) for use in the reconstruction process.
We focus on two TANS schemes for discrete-time stochastic signals:
a greedy method, and a method based on dynamic programming.
We analyze the performances of these schemes by computing (or bounding) their trade-offs between sampling rate and expected reconstruction distortion for autoregressive and Markovian signals.
Simulation results support the analysis of the sampling schemes. We show that
by opportunistically adapting to local signal characteristics TANS may lead to improved power efficiency in some applications.
\end{abstract}


\section{Introduction}\label{sec:introduction}

Sampling is essential in any digital system that interfaces with the
analog world. All else being equal, it is desirable to minimize the
number of samples while maintaining an acceptable reconstruction distortion.
In some applications, minimizing the number of samples can be translated into
having a power-efficient sampling, since the power consumption at an
analog-to-digital converter (ADC) is approximately proportional to its
sampling rate \cite{walden}.
Also, having fewer samples can increase the efficiency of other processing of these measurements. For example, if these samples should be transmitted to another place via a communication channel, having fewer samples will improve power and bandwidth efficiencies.  

One can view \emph{sampling} as a \emph{query} to obtain information from a signal or a function that can only be measured remotely. We get a sample of this function at
an arbitrary time when we \textit{query} for it over a communication
medium.  Here, a portion of the operational cost (e.g., power) is
proportional to the number of samples that we acquire. Hence, again it is desirable to minimize the number of samples taken.

A uniform sampling at the Nyquist rate of the signal may cause some redundant samples, since the global signal bandwidth may not be a good measure of local variations of the signal. Although traditional nonuniform sampling schemes (e.g., \cite{nonunibook,qaisar}) deal with this problem, they have certain limitations. Firstly, they are mostly designed to operate under specific conditions for restrictive signal models (e.g.,  \cite{denis,timevary,timewarp,nonbandunser,vaidyicassp}) and, secondly, sampling times (i.e., time stamps) must be stored/transmitted to be used in the reconstruction process. This may cause power/bandwidth inefficiencies in sampling/communication procedures.   

\begin{figure}
	\centering
    \includegraphics[width=8.5cm]{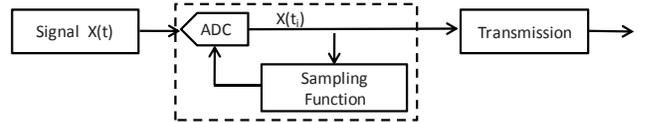}
    \caption{A schematic view of the TANS framework: sampling times are determined by a function of $m$ most recently taken samples. Hence, it is not necessary to save sampling times (time stamps) for use in the reconstruction process.}
    \label{fig:TANS}
  \end{figure}

In this paper, we introduce a new framework for an adaptive nonuniform sampling scheme (see Figure~\ref{fig:TANS}). The key idea of this framework is that \textit{time increments between samples are computed by using a function of previously taken samples}. Therefore, keeping sampling times (time stamps), except initialization times, is not necessary. The function by which sampling time intervals is computed is called the \textit{sampling function}. The aim of this sampling framework is to have a balance between the reconstruction distortion and the average sampling rate. We refer to this sampling framework as \textit{Time-stampless Adaptive Nonuniform Sampling} (TANS).
The TANS concept can be applied on continuous- or discrete-time signals, and the design and analysis can be based on deterministic or stochastic models.

The TANS framework is described in general terms in Section~\ref{sec:framework}.
Section~\ref{sec:problem-setup} then formalizes the problem setup for this
paper, where we focus on discrete-time signals and stochastic models.
A greedy method is developed in Section~\ref{sec:greedy}, and
a method based on dynamic programming (DP) is developed in Section~\ref{sec:dp}.
Simulations results are provided in Section~\ref{sec:simulation},
and some proofs are relegated to Section~\ref{sec:proof}. Section~\ref{sec:conc} concludes the paper.

\section{TANS Framework}\label{sec:framework}

In this section, we introduce the TANS framework and sketch design approaches that will be developed in more detail in later sections. Fix some nonnegative integer $m$ and suppose the $i$th sample of signal $X(t)$ is taken at time $t_i$. We take the $(i+1)$st sample after a time increment of
$$T_{i}=f\left(\left\{(t_j,X(t_j)):i-m+1\leq j\leq i\right\}\right),$$
where $f$ is called the sampling function.
This makes the sampling rate adapt to local characteristics of the signal. 
Since the time increment is a function of the $m$ most recently taken samples, we say the \emph{order} of the sampling function $f$ is $m$.
The sampling is nonuniform except in the trivial cases when $f$ is a constant-valued function (e.g., $m=0$).
Some initialization of the first $m$ sampling times is necessary, but the effect of this initialization on the rate is amortized.

The sampling function is known at the reconstruction side.
Assuming that the \emph{state}
$$\sni=\{(t_j,X(t_j)): i-m+1\leq j\leq i\}$$
is also known at the reconstruction side when reconstructing $X(t)$ on
$[t_i,t_{i+1}]$, there is \emph{no need for the sampling times (time stamps)
to be transmitted}.
These times can be computed by using the sampling function and previously taken samples:
$$t_{i+1}=t_i+f\left(\sni\right).$$
This type of synchronization in an adaptive system without explicit communication is often called backward adaptation~\cite{Gibson:78}.
In a practical setting involving both sampling and quantization,
backward adaptivity requires using the quantized values
to drive the adaptation~\cite{Jayant:73,Gibson:80,OrtegaV:97,GoyalZV:00}.
Here, to maintain focus on sampling rate and adaptation of sampling increments,
we do not explicitly include quantization effects.
Note that while the sampling time selection is causal, the reconstruction method can be causal or non-causal. 

The aim of TANS is to balance between the average sampling rate and the reconstruction distortion. This objective is different from the one considered in change point analysis~\cite{changepoint} or active learning~\cite{active}. There, the objective is to find points of the signal at which statistical behaviors of the signal change, by causal or non-causal sampling, respectively.

Suppose $\Xh(t)$ is the reconstructed signal computed by some reconstruction method. For the case of discrete time and a stochastic signal model, define $d(\sni,T_i)$ as the expected reconstruction distortion over samples from time $t_i+1$ until time $t_{i+1}-1$. That is,
\begin{equation}
d(\sni,T_i)=\mathbb{E}_{\mathcal{X}} \left[\sum_{t=t_i+1}^{t_{i+1}-1} D\big(X(t),\Xh(t)\big)\right], \nonumber 
\end{equation}
where $\mathcal{X}$ is the known probabilistic model of the signal $X(t)$ and $D(X(t),\Xh(t))$ represents the distortion at time $t$.%
\footnote{An analogous formulation for continuous time would replace the sum with an integral over $t \in [t_i,t_{i+1}]$.  Without a stochastic model, a maximum error criterion could be used.}
Note that at times $t_i$ and $t_{i+1}$ the reconstruction distortion is zero since exact sample values are known at these times. In realistic cases and for a given state $\sni$, $d(\sni,T_i)$ is an increasing function with respect to $T_i$, because the greater the next sampling step, the greater the reconstruction distortion. On the other hand, the greater the next sampling step, the larger the rate benefit. Hence, a rate penalty can be defined as $a(\sni,T_i)=\Frac{\rho}{f(\sni)}=\Frac{\rho}{T_i}$, where $\rho$ is a rate award parameter. We define the cost of each sampling state as the sum of the expected reconstruction distortion and the rate penalty, that is, $c(\sni,T_i)=d(\sni,T_i)+a(\sni,T_i)$. The overall cost of the sampling process is the sum of different sampling state costs, that is, $\sum_{i}c(\sni,T_i)$.    

Finding an appropriate sampling function for TANS depends on requirements such as average sampling rate, maximum distortion, etc.  
In this paper, we investigate two general approaches to computing appropriate sampling functions for given sampling setups: \textit{greedy methods} and \textit{dynamic programming (DP) methods}. 

\begin{figure}
	\centering
    \includegraphics[width=8.5cm]{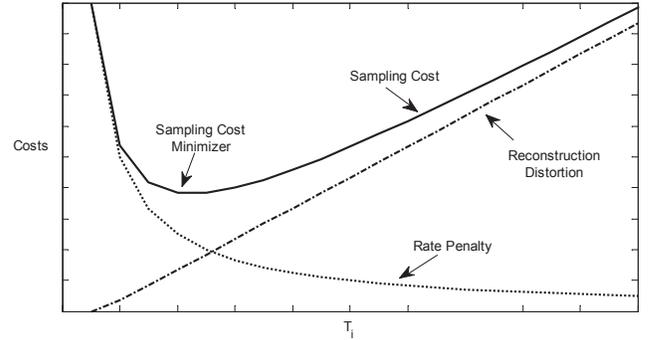}
    \caption{Demonstrating the behavior of different parts of sampling state cost $c(\sni,T_i)=d(\sni,T_i)+a(\sni,T_i)$, where $d(\sni,T_i)$ is the reconstruction distortion and $a(\sni,T_i)$ is the rate penalty function.}
    \label{fig:greedy-cost}
  \end{figure}

In greedy methods, a sampling function at state $\sni$ chooses the next sampling increment $T_i$ to minimize the sampling state cost $c(\sni,T_i)$. As depicted in Figure~\ref{fig:greedy-cost}, for any given state $\sni$, $d(\sni,T_i)$ is an increasing function of $T_i$, while $a(\sni,T_i)$ is a decreasing function. Therefore, there is a trade-off between the sampling rate and the expected reconstruction distortion. A greedy method balances this trade-off by choosing the state cost minimizer as the next sampling increment. In certain cases, greedy sampling schemes can perform closely to an optimal scheme.

Since $c(\sni,T_i)$ depends on the current sampling state $\sni$, a greedy sampling function does not take into account characteristics of the next sampling state.  Intuitively, the larger the sampling increment $T_i$ at the sampling state $\sni$, the lower the \textit{quality} of the next sampling state. Hence, in general, greedy methods are not optimal sampling schemes considering the overall sampling cost $\sum_{i}c(\sni,T_i)$ as a comparison measure. We consider effects of the next sampling states' \textit{quality} in DP methods. We show that an exact Bellman-Ford equation (BFE) can be written and solved for some sampling setups. For those cases, the solution of BFE provides an optimal sampling function which minimizes the overall sampling cost. In cases where solving the BFE is not practically feasible (because the number of possible sampling states is large or exact sampling states are not known), we propose sampling functions based on approximate dynamic programming (ADP) algorithms. Sampling functions derived by greedy methods can be used in ADP-based sampling methods. In fact, greedy methods can be viewed as DP-based methods with a unit time horizon.

In this paper, we consider two examples of stochastic signals: stationary signals and Markovian signals. Unlike stationary signals, Markovian processes have sudden changes in their statistical properties based on an underlying hidden Markov chain. Note that the general sampling framework can be applied on other signal models.


\section{Problem Setup, Signal Models and Background Results}\label{sec:problem-setup}

In this section, we first present the problem setup of TANS\@. Then, we introduce signal models considered in this paper. At the end of this section, a \textit{generalized linear prediction filter} is proposed, which linearly predicts the future samples of a stationary process by using a set of nonuniform samples from its past. 

\subsection{Problem Setup}\label{subsec:methods}

Consider a discrete-time signal $X(t)$. TANS with order $m$ is used to take samples from this signal. Hence, the next sampling increment at time $t_i$ is a function of $m$ most recently taken samples at that time (i.e., $T_i=f(\sni)$). The set of samples taken using the sampling function $f(\cdot)$ is denoted by $\Gamma_{X(t)}^{f}$. A function $g(\cdot)$ is used to reconstruct the original signal from its samples, that is, $\Xh(t)=g(\Gamma_{X(t)}^{f})$. The reconstruction error signal is $e(t)=X(t)-\Xh(t)$. 
The overall sampling cost is the sum of sampling state costs over different states, that is, $\ctot(X(t),f,g)=\sum_{i} c(\sni,T_i)$. 

A system optimization problem under the TANS framework can be stated as follows:
\begin{defn}[Optimal TANS sampling problem]\label{def:sampling-problem}
For a class of signals $X(t)$, a given reconstruction function $g(\cdot)$, and an order $m$, a sampling function $f^{*}(\cdot)$ is desired to minimize the overall expected sampling cost:
\begin{equation}
f^{*}=\arginf_{f} \ctot(f,g).
\end{equation}
The resulting cost is denoted $\ctot^* = \ctot(f^{*},g)$. 
\end{defn}


\subsection{Signal Models}\label{subsec:signal-models}

\begin{figure}
	\centering
    \includegraphics[width=6cm]{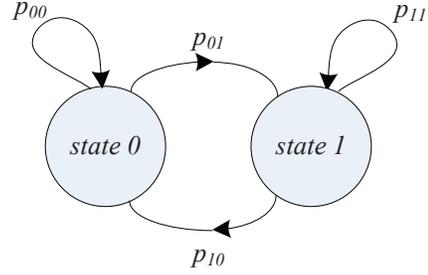}
    \caption{A hidden Markov chain considered in Markovian signal model of equation \eqref{eq:signal-model-2}.}
    \label{fig:MC}
  \end{figure}

In this paper, we consider the following signal models:

\begin{itemize}
\item Case 1: an autoregressive signal with memory of one (i.e., AR(1)):
\begin{equation}\label{eq:signal-model-1}
X(t+1)=\alpha X(t)+ Z(t+1),
\end{equation}

where $Z(t+1)$ is a Gaussian noise with zero mean. If the power of the signal is assumed to be one, the noise variance is $1-\alpha^2$. 

\item Case 2: a Markovian signal:
\begin{equation}\label{eq:signal-model-2}
X(t+1)=\alpha_{\theta_t} X(t)+ Z_{\theta_t}(t+1),
\end{equation}

where $\theta_t$ represents the state of a hidden Markov chain (MC) with state transition probabilities depicted in Figure~\ref{fig:MC}. At time $t$, if the MC is at state $0$, $\theta_t=0$; otherwise, $\theta_t=1$. Depending on the value of $\theta_t$, the signal is generated by a first-order AR model with parameter $\alpha_{\theta_t}$ and the noise variance $1-\alpha_{\theta_t}^2$. Note that, in this model, unlike the previous case, the coefficient of the AR model has a sudden change in time depending on the state of the underlying hidden Markov chain.
\end{itemize}

\subsection{Generalized Linear Prediction Filter}\label{subsec:generalized-linear-predict}

Suppose $X(t)$ is a stationary signal. Assume we have $m$ samples of $X(t)$ at times $\{t_{i-m+1},\ldots,t_{i}\}$. Our aim is to linearly predict $X(t)$, where $t=t_i+T$ for some $T\geq 1$, by using these known samples so that the expected mean square error is minimized (MMSE predictor).

 Consider $\Xh(t)$ as a predicted value of $X(t)$ by using these $m$ sample values. The prediction error is $e(t)=X(t)-\Xh(t)$. Define $\tau_k=t-t_{i-k}$ for $0\leq k\leq m-1$. We want to find optimal linear prediction weights $w_{\tau_0},\ldots,w_{\tau_{m-1}}$ so that the prediction error power is minimized: 

\begin{align}\label{eq:opt-gen-linear-filter}
\min_{w_{\tau_0},\ldots,w_{\tau_{m-1}}}\quad& E[|e(t)|^2] \\
\mbox{subject to}\quad& \Xh(t)=\sum_{k=0}^{m-1} w_{{\tau_k}} X(t_{i-k}).\nonumber 
\end{align}

A solution of this linear optimization is referred as $w_{\tau_i}^*$, for $0 \leq i\leq m-1$. Note that, unlike a regular linear prediction filter (\cite{Haykin:96}), a generalized linear prediction filter predicts $X(t)$ by using a set of nonuniform samples.

The auto-correlation function of $X(t)$ can be written as
\begin{equation}\label{eq:auto}
r(i)=E\big[X(t)X^{c}(t-i)\big],
\end{equation}
where $X^{c}(t)$ represents the complex conjugate of $X(t)$. In this paper, we deal with real signals.  

To simplify notations, we define the following matrices: 
\begin{eqnarray}\label{eq:matrix}
\btau &=&[\tau_0,\ldots,\tau_{m-1}]^T\\
\bp&=&\big[r(-\tau_0),\ldots, r(-\tau_{m-1})\big]^{T}\nonumber\\
\bw_{\btau}^{*}&=&\big[w_{\tau_0}^{*}, ..., w_{\tau_{m-1}}^{*}\big]^T.\nonumber
\end{eqnarray}
Also, a $m\times m$ auto-correlation matrix $\bR$ is defined whose component in the $i$th row and $j$th column is $r(\tau_i-\tau_j)$.

The following theorem provides optimal weights for the generalized linear prediction filter:
\begin{thm}\label{thm:weights}
\begin{equation}
\bp=\bR \bw_{\btau}^{*}.\nonumber 
\end{equation}
\end{thm}
\begin{IEEEproof}
See Section~\ref{sec:proof-thm:weights}.
\end{IEEEproof}

For $X(t)$ with zero mean, the variance of the prediction error is defined as follows:
\begin{eqnarray}
\sigma_{e^{*}}^2(\sni,T)= E[|e^{*}(t)|^2].\nonumber
\end{eqnarray}

The following theorem provides a way to compute the variance of the prediction error for a generalized linear prediction filter:

\begin{thm}\label{thm:gen-linear}
$\sigma_{e^{*}}^2(\sni,T)=r(0)-\bp^{T} \bw_{\btau}^{*}.$
\end{thm}
\begin{IEEEproof}
See Section~\ref{sec:proof-thm:gen-linear}.
\end{IEEEproof}


\section{Greedy TANS}\label{sec:greedy}
In this section, we investigate greedy sampling functions various signal models. In Section~\ref{sec:dp}, we evaluate performance of these schemes and compare them with some other schemes, including uniform sampling setups. In all of these sampling schemes, the reconstruction function is assumed to be a generalized linear prediction filter, introduced in Section~\ref{subsec:generalized-linear-predict}. Note that it is a causal reconstruction function. 

In greedy methods, a sampling function is computed as follows:
\begin{align}\label{eq:greedy}
T_i=\argmin_{T}\quad& c(\sni,T) \\
\mbox{subject to} \quad& T\geq 1,\nonumber
\end{align}
where $f(\sni)=T_i$. The sampling function of a greedy method depends on the current sampling state and does not take into account characteristics of the next sampling states. Intuitively, the larger the sampling increment $T_i$ at the sampling state $\sni$, the lower the \textit{quality} of the next sampling state. Therefore, we have a trade-off between the sampling rate award of the current state and the sampling cost of the next state (Figure \ref{fig:greedy-cost}). Hence, greedy methods usually are not optimal solutions of a TANS sampling problem presented in Definition~\ref{def:sampling-problem}. However, these greedy solutions can be used to approximate optimal solutions, which may have high computational complexity.

Here, we investigate the greedy sampling function of \eqref{eq:greedy} for two signal classes described in Section \ref{subsec:signal-models}. We analyze their sampling rate versus the expected reconstruction distortion in Theorems~\ref{thm:stat} and~\ref{thm:greedy-markovian}. Simulation results of proposed schemes are shown in Section~\ref{sec:simulation}.

\subsection{Greedy TANS for Autoregressive Signals}\label{subsec:greedy-AR}

In this section, we consider greedy sampling functions for an AR(1) signal model described in \eqref{eq:signal-model-1}.

Suppose $f_{\rm greedy}^*(\sni)$ is an optimal greedy sampling function. Also, suppose $\dtot^*$ is the expected reconstruction distortion per sample corresponding to this sampling function. 
The following theorem introduces an optimal greedy sampling function and its expected reconstruction distortion per sample.

\begin{thm}\label{thm:stat}
For an AR(1) signal with parameter $\alpha$ described in \eqref{eq:signal-model-1}, over a large enough time interval $[0,\Ttot]$, an optimal greedy sampling function is
\begin{eqnarray}\label{eq:stat-optimal}
f_{\rm greedy}^*(\sni) & = & T^*,\\
\dtot^* & = & \frac{1}{T^*} \sum_{j=1}^{T^*-1} 1-\alpha^{2j},\nonumber
\end{eqnarray}
where $T^*=\argmin_{T} \sum_{j=1}^{T-1} (1-\alpha^{2j})+\Frac{\rho}{T}.$
\end{thm}
\begin{IEEEproof}
See Section~\ref{sec:proof-thm:stat}.
\end{IEEEproof}

Note that, for this signal model, an optimal greedy sampling function yields uniform sampling with the sampling rate $\Frac{1}{T^{*}}$, which performs closely to an optimal sampling scheme. In the following corollary, we present a formula to compute $T^*$.

\begin{cor}\label{cor:stat}
Suppose $\Troot$ is a solution of the following equation:
\begin{equation}\label{eq:root}
(1-\alpha^{2T})-\frac{\rho}{T(T+1)}=0
\end{equation}
where $0<\alpha<1$ and $\rho>0$. If $(1-\alpha^2)<\frac{\rho}{2}$, then
\begin{itemize}
\item[(i)]  $\Troot$ is unique; and
\item[(ii)] $T^*=\lfloor\Troot\rfloor$ or $T^*=\lfloor\Troot\rfloor+1$.
\end{itemize}
\end{cor}
\begin{IEEEproof}
See Section~\ref{sec:proof-cor:stat}.
\end{IEEEproof}

We will validate this by simulation in Section~\ref{sec:simulation}. Also, arguments of Theorem~\ref{thm:stat} can be extended for a general stationary signal. 

For an AR signal, since statistical properties of the signal do not vary in time, there is no rate adaption with respect to sample values. This is not the case for Markovian signals. We investigate the greedy TANS framework for Markovian signals in the next section; a rate adaption with respect to sample values would be helpful to minimize the sampling cost and leads to a nonuniform sampling scheme. 

\subsection{Greedy TANS for Markovian Signals}\label{subsec:greedy-markovian}

Consider a Markovian signal described by \eqref{eq:signal-model-2}, where $\theta_t$ represents the state of a hidden underlying Markov chain depicted in Figure~\ref{fig:MC}. In this section, for simplicity we assume the MC is symmetric (i.e., $p_{01}=p_{10}$). However, all arguments can be extended for a general MC\@. We also assume that $\alpha_0$ and $\alpha_1$ are known.  However, the state of the Markov chain (i.e., $\theta_t$) is unknown and needed to be estimated by using the taken samples. We use a generalized linear prediction filter for the reconstruction. 
Note that if $\alpha_0$ and $\alpha_1$ are also unknowns, one can learn these parameters at the beginning of the process by taking more samples. Then, our proposed scheme can be applied for the rest of the process.

Extending the previous notation, define $\theta_{\sni}$ as the state of the MC during the sampling state $\sni$. If during $\sni$ the MC state stays at zero, $\theta_{\sni}=0$. Similarly, if the MC state stays at one, $\theta_{\sni}=1$. Otherwise, if there is an MC transition within this sampling state, $\theta_{\sni}=2$. We assume that $\theta_{\sni}$ is unknown and needs to be estimated by using the taken samples. The estimated value of $\theta_{\sni}$ is referred by $\htsni$. The error probability of this estimation is referred by $P_e(\sni)=Pr(\htsni\neq\theta_{\sni})$. 

\begin{alg}\label{def:greedy-markovian}
A greedy sampling function for the considered Markovian signal has the following steps:

\begin{itemize}
\item Step i,0: Compute $\htsni$ and $P_e(\sni)$.
\item Step i,1: Compute $T_i=\argmin_{T} c(\sni,T|\htsni)$, where $c(\sni,T|\htsni)$ is the sampling state cost given $\htsni$ (see \eqref{eq:cost-markov-apx} and \eqref{eq:cost-markov2}).
\item Step i,2: Take a sample at time $t_i+T_i$. 
\item Step i,3: Compute $S_{t_i+1}$. Repeat.
\end{itemize}
\end{alg}

For simplicity, we assume that the sampling increment $T$ is small enough that the probability of having more than one MC transition is negligible. In other words, we have the following assumptions.

\begin{asm}\label{asm:1}
\begin{eqnarray}\label{eq:conditions-markovian}
\max_{i} T_i & \leq & T_{up}, \nonumber\\
p_{00}^{(T_{up}-1)} & \gg & \frac{1}{2},\nonumber\\
p_{11}^{(T_{up}-1)} & \gg & \frac{1}{2}.\nonumber
\end{eqnarray}
\end{asm}

If MC transition probabilities $p_{01}$ and $p_{10}$ are small enough, this assumption is reasonable. Under this assumption, the sampling state cost, $c(\sni,T)$, can be conditioned on the value of $\htsni$ as follows:
\begin{eqnarray}\label{eq:cost-markov1}
\lefteqn{c(\sni,T \, | \, \htsni=0)} \nonumber \\
& = & Pr(\htsni=\theta_{\sni})\Big\{\sum_{j=0}^{T-1} p_{00}^j p_{01} \sum_{l=1}^{j}(1-\alpha_0^{2l})\nonumber\\
& & + (T-1-j)\maxervar \Big\}\nonumber\\
& & + Pr(\htsni\neq\theta_{\sni})\maxervar+\frac{\rho}{T} \nonumber \\
& = & (1-P_e(\sni))\Big\{\sum_{j=0}^{T-1} p_{00}^j p_{01} \sum_{l=1}^{j}(1-\alpha_0^{2l}) \nonumber\\
& & + (T-1-j)\maxervar \Big\}\nonumber\\
& & + P_e(\sni) (T-1)\maxervar+\frac{\rho}{T},
\end{eqnarray}
which says that, if the estimation is correct (with probability $1-P_e(\sni)$) and an MC transition happens at time $t_i+j+1$ (with probability $p_{00}^j p_{01}$), the sampling state cost is $\sum_{l=1}^{j}1-\alpha_0^{2l}+(T-1-j)\maxervar$, where $\maxervar$ is the maximum prediction error variance (in this example, $\maxervar=1$). If the estimation process fails (with probability $P_e(\sni)$), a maximum prediction error variance $\maxervar$ occurs. By using the assumption $p_{00}^{T_{up}} \gg 1/2$, equation \eqref{eq:cost-markov1} can be simplified as follows: 
\begin{eqnarray}\label{eq:cost-markov-apx}
\lefteqn{ c(\sni,T \, | \, \htsni=0) } \nonumber \\
& \approx & (1-P_e(\sni)) \sum_{\ell=1}^{T-1} (1-\alpha^{2\ell}) \nonumber\\
& & +P_e(\sni) (T-1)\maxervar+\frac{\rho}{T}.
\end{eqnarray}

The sampling state cost function conditioned on $\htsni=1$ (i.e., $c(\sni,T \, | \, \htsni=1)$) can be written similarly.

Finally, for the case $\htsni=2$, we assume that the prediction variance is the maximum prediction error variance $\maxervar$:
\begin{equation}\label{eq:cost-markov2}
c(\sni,T \, | \, \htsni=2)=(T-1)\maxervar+\frac{\rho}{T}.
\end{equation}

We analyze the performance of the proposed greedy sampling scheme in the following theorems. We derive upper and lower bounds for average sampling rate and expected reconstruction distortion per sample of the proposed sampling scheme. Simulation results for this sampling scheme are presented in Section \ref{sec:simulation}, which support the derived analytical bounds.

Before presenting theorems, we introduce some notations. Suppose that, for all $\sni$, $P_e^{\rm low}\leq P_e(\sni)\leq P_e^{\rm up}$. By considering an upper bound on $P_e(\sni)$, we define
\begin{align}
T_0^{\rm low}=\argmin_{T}\quad &(1-P_e^{\rm up}) \sum_{l=1}^{T-1} (1-\alpha_0^{2l}) \nonumber\\
&+P_e^{\rm up} (T-1)\maxervar+\frac{\rho}{T}.
\end{align}
$T_0^{\rm up}$ is defined similarly by considering a lower bound on $P_e(\sni)$. Analogously, $T_1^{\rm up}$ and $T_1^{\rm low}$ can be defined.

Also, $d_0^{\rm up}$, an upper bound on the expected reconstruction distortion per sample given $\htsni=0$ is defined as follows:
\begin{equation}
d_0^{\rm up}=\frac{1}{T_0^{\rm up}}\big\{(1-P_e^{\rm up}) \sum_{\ell=1}^{T_0^{up}-1} (1-\alpha_0^{2\ell}) +P_e^{up} (T_0^{up}-1)\maxervar\big\}.\nonumber
\end{equation}
Quantities $d_0^{low}$, $d_1^{up}$ and $d_1^{low}$ are defined similarly. 

The following theorem provides analytical upper and lower bounds on the average sampling rate and the expected reconstruction distortion of the greedy sampling scheme introduced in Algorithm~\ref{def:greedy-markovian}.  

\begin{thm}\label{thm:greedy-markovian}
Consider a Markovian signal defined in \eqref{eq:signal-model-2} over a large enough time interval $[0,\Ttot]$. Under Assumption~\ref{asm:1}, an achievable rate-distortion pair $(R,D)$ of the greedy sampling scheme of Algorithm~\ref{def:greedy-markovian} can be bounded as follows:
\begin{eqnarray}
\frac{1}{2T_0^{\rm up}}+\frac{1}{2T_1^{\rm up}}\leq &R&\leq \frac{1}{2T_0^{\rm low}}+\frac{1}{2T_1^{\rm low}}\\
\frac{d_0^{\rm low}}{2}+\frac{d_1^{\rm low}}{2}\leq &D&\leq \frac{d_0^{\rm up}}{2}+\frac{d_1^{\rm up}}{2}.
\end{eqnarray}
\end{thm}
\begin{IEEEproof}
See Section~\ref{sec:proof-thm:greedy-markovian}.
\end{IEEEproof}

Similarly to Corollary~\ref{cor:stat}, $T_0^{\rm low}$, $T_0^{\rm up}$, $T_1^{\rm low}$ and $T_1^{\rm up}$ can be calculated by finding roots of some equations. For example: 
\begin{cor}\label{cor:greedy-markovian}
Suppose $\Troot$ is a solution of the following equation:
\begin{equation}
(1-P_e^{\rm up})(1-\alpha_0^{2T})+P_e^{\rm up}(T-1)\maxervar-\frac{\rho}{T(T+1)}=0
\end{equation}
where $0<\alpha<1$ and $\rho>0$. If $(1-P_e^{\rm up})(1-\alpha_0^2)<\frac{\rho}{2}$, then, 
\begin{itemize}
\item[(i)]  $\Troot$ is unique; and
\item[(ii)] $T_0^{\rm low}=\lfloor\Troot\rfloor$ or $T_0^{\rm low}=\lfloor\Troot\rfloor+1$.
\end{itemize}
\end{cor}

A similar corollary can be stated for $T_0^{\rm up}$, $T_1^{\rm low}$ and $T_1^{\rm up}$. 

In Theorem~\ref{thm:greedy-markovian}, the performance of the proposed sampling scheme (i.e., its average sampling rate and the expected reconstruction distortion) is bounded. However, it is insightful to compare its performance to a genie-aided sampling scheme where the state of the underlying Markov chain is known. For a genie-aided scheme, $P_e^{\rm low}=P_e^{\rm up}=0$, and therefore upper and lower bounds of Theorem~\ref{thm:greedy-markovian} match:

\begin{cor}\label{cor:genie}
For a genie-aided sampling scheme, the following rate-distortion pair is achievable:
\begin{eqnarray}
R&=&\frac{1}{2T_0^{\rm genie}}+\frac{1}{2T_1^{\rm genie}}, \\
D&=&\frac{d_0^{\rm genie}}{2}+\frac{d_1^{\rm genie}}{2},
\end{eqnarray}
where
\begin{align}
T_0^{\rm genie}&=\argmin_{T}  \sum_{\ell=1}^{T-1} (1-\alpha_0^{2\ell}) +\frac{\rho}{T},\nonumber\\
d_0^{\rm genie}&=\frac{1}{T_0^{\rm genie}}\big\{ \sum_{\ell=1}^{T_0^{\rm genie}-1} (1-\alpha_0^{2\ell})\big\},\nonumber 
\end{align}
and
$T_1^{\rm genie}$ and $d_1^{\rm genie}$ are defined similarly.
\end{cor}

The proof of Corollary~\ref{cor:greedy-markovian} is similar to the one of Corollary~\ref{cor:stat}. Also, Corollary~\ref{cor:genie} can be derived by using $P_e^{\rm low}=P_e^{\rm up}=0$ in Theorem~\ref{thm:greedy-markovian}.

In the first step of this sampling function, we need to estimate $\theta_{\sni}$ by using $m$ most recently taken samples of the signal and compute its probability of error $P_e(\htsni)$. In the following, we give an example of such a scheme.

\begin{ex}
Suppose $m=2$ and we  use a maximum likelihood estimator.
If $\theta_{\sni}=0$, the probability distribution of $X(t_i)$ is $\cN(\alpha_0^{T_{i-1}} X(t_{i-1}),1-\alpha_0^{2T_{i-1}})$, where $\cN(\mu,\sigma^2)$ represents a Gaussian distribution with mean $\mu$ and variance $\sigma^2$. The prior probability of this event is $p_{00}^{T_{i-1}}$. Similarly, if $\theta_{\sni}=1$, $X(t_i)$ is distributed by $\cN\big(\alpha_1^{T_{i-1}} X(t_{i-1}),1-\alpha_1^{2T_{i-1}}\big)$. The prior probability of this event is $p_{11}^{T_{i-1}}$. Otherwise, the distribution of $X(t_i)$ is $\cN(\alpha_0^{j}\alpha_1^{T_{i-1}-j} X(t_{i-1}),1-\alpha_0^{2j}\alpha_1^{2(T_{i-1}-j}))$ with the prior probability ${T_{i-1}\choose j}p_{00}^j p_{11}^{T_{i-1}-j}$, for $1\leq j\leq T_{i-1}-1$. Each of these events corresponds to the case $\theta_{\sni}=2$ (i.e., there is a transition within the sampling state.). Therefore, by having the observed value of $X(t_i)$ and using these distributions, a maximum likelihood estimator can estimate $\htsni$ and compute its error probability. 
\end{ex}

\textit{Remarks:}
\begin{enumerate}
\item If the state space is not large, computations can be performed off-line and results can be used in the sampling function.
\item When the order of the sampling function is large and/or autocorrelation coefficients change continuously, a maximum likelihood estimator may not be practically interesting. In these cases, we can use previously taken samples within a window of size $W$ from the last sample (i.e., all taken samples from time $t_i-W+1$ to the time $t_i$) to update or estimate autocorrelation coefficients to use in the sampling function. The quality of this estimation process depends on the window size $W$, the variation rate of autocorrelation coefficients, and the technique used. Two possible methods for estimating autocorrelation coefficients are as follows:
\begin{itemize}
\item A gradient-based method. Suppose at the sampling state $S_{t_{i-1}}$, the set of estimated autocorrelation coefficients is $\{\rh(j):j\geq 1\}$. By taking a sample at time $n_i$, these coefficients are updated as follows:
\begin{equation}
\rh(j):=\rh(j)+\gamma (X(t_i)X(t_i-j)-\rh(j))
\end{equation}
for all possible $j$'s, where $:=$ represents an update sign, and $\gamma>0$ is a gradient step size. This gradient-based update method can be useful when $W$ is not large.
\item A window-based method. If the window size $W$ is large and there are enough known samples within the window, an empirical value for each autocorrelation coefficient can be computed.   
\end{itemize}
The estimated autocorrelation coefficients $\{\rh(j):j\geq 0\}$ can be used in the generalized linear prediction filter in order to design a sampling function similar to the one of Algorithm~\ref{def:greedy-markovian}.
\end{enumerate}


\section{Dynamic Programming-Based TANS}\label{sec:dp}

In greedy TANS, sampling functions are derived based on minimizing the sampling cost at each sampling state. Hence, it does not take into account the \textit{quality} of next sampling states. Intuitively, the larger the sampling increment at the sampling state $\sni$, the lower the quality of the next sampling state. Therefore, in general, greedy methods may not provide optimal sampling functions with respect to the overall sampling cost. 

We consider \textit{quality} of next sampling states in dynamic programming-based TANS methods.
For the sampling state $\sni$, a cost-to-go function $J_{f}(\sni)$ is defined as follows:
\begin{eqnarray}
J_f(\sni)&=&c(\sni,T_i)+\beta c(S_{t_{i+1}},T_{i+1})+\cdots\\
&=&\sum_{j\geq i} \beta ^{j-i} c(S_{t_j},T_j).\nonumber
\end{eqnarray}
In this setup, $0<\beta<1$ is called a \emph{discount factor}. An interpretation of this factor is that the cost of the next sampling state is less important for the current state policy by a factor of $\beta$. Another interpretation of this factor is that the process may be ended at each sampling state with probability $1-\beta$.
An optimal cost-to-go function $J(\sni)$ of the sampling state $\sni$ is defined as follows:
\begin{equation}
J(\sni)=\inf_{f} J_f(\sni).
\end{equation}

A Bellman-Ford equation (BFE) can be written for this problem by using cost-to-go functions of different sampling states as follows:
\begin{equation}\label{eq:BFE}
J(\sni)=\inf_{T_i} \big(c(\sni,T_i)+\beta E[J(S_{t_{i+1}})]\big).
\end{equation}
A solution of this Bellman-Ford equation (BFE) is an optimal solution for the sampling problem presented in Definition~\ref{def:sampling-problem} when the reconstruction function is causal. We investigate this problem for various signal models and sampling setups in this section. For some cases where the number of sampling states is not large, an optimal sampling function can be derived. In other cases where finding this optimal solution is computationally difficult, we propose sampling functions based on approximate dynamic programming (ADP) algorithms. We define a \textit{quality} function $q(\sni)$ for each sampling state $\sni$. A greedy solution is used to define this quality function. Then, a sampling function can be computed as follows:
\begin{align}\label{eq:adp}
T_i=\arginf_{T}\quad& c(\sni,T)+\beta E[q(S_{t_{i+1}})] \\
\mbox{subject to}\quad& T\geq 1.\nonumber
\end{align}
In this setup, we consider quality effects of just one sampling state ahead. Also, note that the solution of the BFE is optimal if the reconstruction function $g(\cdot)$ is causal. It is a necessary assumption of the dynamic programming setup to have a separation of sampling state costs at different stages.


\subsection{An Online Source Coding Scheme Based on TANS}\label{subsec:online-source-coding} 

In this section, we consider a sampling problem where an exact DP-based solution can be derived as a solution of the BFE \eqref{eq:BFE}. Consider a Markovian signal with an underlying hidden Markov chain depicted in Figure~\ref{fig:MC}. For simplicity, suppose $p_{01}=\eps_{0} \ll 1$ and $p_{10}=\eps_{1} \ll 1$. Other regimes of transition probabilities can be analyzed similarly. Suppose $X(t)$ is a binary signal generated by this underlying hidden Markov chain so that, at state $0$, $X(t)=0$, and at state $1$, $X(t)=1$. For the reconstruction, we use a causal function that selects a most-probable binary sequence to fill missing places. Hamming distance is used as an error measure. We call this problem an \textit{online source coding} problem since it does not have a compression delay as in an unconstrained block source coding scheme. In this problem, TANS can provide some compression gain without having any delay. This problem can be extended to Markov chains with more states and different regimes of transition probabilities. 

In this sampling problem, there are two different sampling states: If $X(t_i)=0$, then, $\sni=0$; otherwise, $\sni=1$. Suppose $f(\cdot)$ is the sampling function (i.e., $T_i=f(\sni)$.). Suppose $f(\sni=0)=T_0$ and $f(\sni=1)=T_1$. Hence, $T_0$ and $T_1$ represent sampling steps at different sampling states.       

Since $\eps_0,\eps_1 \ll 1$, the reconstruction method would choose all-0 and all-1 sequences when $\sni=0$ and $\sni=1$, respectively. Suppose $\sni=0$. Hence, the next sample is taken at time $t_i+T_0$. An error happens if there is one or more Markov chain transitions over the time interval $[t_i+1,t_i+T_0-1]$. To simplify our analysis, we only consider first-order error terms (i.e., we assume at most one transition happens over a time interval $[t_i+1,t_i+T_0-1]$). To have this simplifying assumption, we need to restrict the sampling increments such that $\max T_0,T_1 \ll \min\{\frac{1}{\eps_0},\frac{1}{\eps_1}\}$. Therefore, the probability of having more than one Markov chain transition over a time interval of a length $T_0$ or $T_1$ is negligible. 

By considering Hamming distance as a distortion measure, the sampling state cost at $\sni=0$ can be written as follows:
\begin{equation}
c(\sni,T_0)=\sum_{j=1}^{T_0-1} (1-\eps_0)^{j-1}\eps_0 (T_0-j)+\frac{\rho}{T_0}.
\end{equation}
Note that, $(1-\eps_0)^{j-1}\eps_0$ is the probability of not having a transition over the first $j-1$ samples of the sampling interval and having a transition at the $j$th sample. Hence, the considered reconstruction method makes errors on samples from time $t_i+j$ to $t_i+T_0-1$, which corresponds to a Hamming distance of $T_0-j$. A similar argument can be made for the case of $\sni=1$.

By considering these sampling state cost functions, the BFE can be written as follows:
\begin{eqnarray}\label{eq:bfe-source-coding}
J(\sni=0)&=&\min_{T_0,T_1} \big[c(\sni=0,T_0)\nonumber\\
& & + \beta (1-\eps_0)^{T_0} J(\sni=0)\nonumber\\
& & + \beta (1-(1-\eps_0)^{T_0})J(\sni=1) \big],\nonumber\\
J(\sni=1)&=&\min_{T_0,T_1} \big[c(\sni=1,T_1)\nonumber\\
& & + \beta (1-\eps_1)^{T_1} J(\sni=1)\nonumber\\
& & + \beta (1-(1-\eps_1)^{T_1})J(\sni=0) \big].\nonumber
\end{eqnarray}
Since these BFEs have only two variables, various numerical and analytical methods can be applied to find their solution, which in turn corresponds to an optimal sampling scheme in TANS (e.g., see \cite{dpbook}). Simulation results for this sampling scheme are given in Section~\ref{sec:simulation}.


\subsection{Approximate Dynamic Programming Methods} \label{subsec:adp-tans}
Finding a solution of the BFE of a DP-based sampling function may not be practically feasible if the sampling state space is large or is unknown (or is known partially). In these cases, approximate dynamic programming (ADP) algorithms can provide suboptimal solutions with a reasonable computation complexity. In this section, we investigate an ADP-based sampling function of TANS introduced in \eqref{eq:adp}. To use this approximate algorithm, for any sampling state $\sni$, a \textit{quality} measure $q(\sni)$ is assigned. We use a greedy sampling solution to define this quality function. 

Consider a Markovian signal $X(t)$ described in Section~\ref{subsec:signal-models}. A greedy sampling function for this signal is introduced in Algorithm~\ref{def:greedy-markovian}. We refer to this greedy sampling function as $f^{\rm greedy}(\cdot)$, where $T_i^{\rm greedy}=f^{\rm greedy}(\sni)$.

A \textit{quality} function of each state is defined as
\begin{equation}\label{eq:quality}
q(\sni)=\gamma T_i^{\rm greedy}
\end{equation}
where $\gamma$ is a scaling parameter. Intuitively, the larger the greedy sampling step, the higher the quality of the sampling state. 
Therefore, an ADP-based sampling function of TANS can be derived by using the optimization setup of \eqref{eq:adp}. 

For computing the expected quality of the next sampling state, transition probabilities among different sampling states should be known. Suppose, at a sampling state $\sni$, the next sample is taken after a time interval $T_i$. The value of the sample $X(t_i+T_i)$ is a random variable with a mean $\Xh(t_i+T_i)$, which can be computed by a generalized linear prediction filter. The probability distribution of $X(t_i+T_i)$ determines the probability distribution of the next sampling state $S_{t_{i+1}}$. Therefore, the expected value of the quality function of the next state can be computed by using this probability distribution. However, to simplify this sampling function further, one may approximate this expected quality by the quality of the most probable next state, which has a sample value of $\Xh(t_i+T_i)$ at time $t_{i+1}$. We call this state $\Sh_{t_{i+1}}$. Therefore, a more simplified sampling function based on an ADP algorithm can be written as
\begin{eqnarray}\label{eq:adp2}
T_i&=&\argmin_{T}\quad c(\sni,T)+\beta q(\Sh_{t_{i+1}}).\nonumber
\end{eqnarray}

For any given state $\sni$ and $T_i=T$, $q(\Sh_{t_{i+1}})$ can be  computed. Intuitively, the term $\beta q(\Sh_{t_{i+1}})$ is a correction term for the greedy solution considering the quality of the next sampling state. In this scheme, the quality of only one future sampling state is considered. However, one can extend this algorithm to consider the quality of more than one future sampling state. 
\begin{alg}\label{alg:adp-tans}
An approximate dynamic programming-based sampling function for the Markovian signal of \eqref{eq:signal-model-2} can be summarized as follows:
\begin{itemize}
\item Step i,0: Compute $\htsni$, $P_e(\sni)$ and $q(\Sh_{t_{i+1}})$.
\item Step i,1: Compute $T_i=\min{c(\sni,T|\htsni)}+\beta q(\Sh_{t_{i+1}})$.
\item Step i,2: Take a sample at time $t_i+T_i$. 
\item Step i,3: Compute $S_{t_i+1}$. Repeat.
\end{itemize}
\end{alg}

Simulation results for this sampling procedure are presented in Section~\ref{sec:simulation}.

\section{Simulation Results}\label{sec:simulation}
In this section, we evaluate the performance of the proposed sampling schemes by simulations and compare their performance against uniform sampling. In
uniform sampling, the sampling rate is always in the form of $R = 1/T_{\rm uni}$,
, where $T_{\rm uni}$ is a positive integer. To be able to
compare the performance of different methods with uniform
sampling at different rates, we modify the uniform
sampling setup to capture all possible sampling rates. To do
this, for a given rate $R = 1/T_{\rm uni}$ where $T_{\rm uni}$ is not an integer
number, the $i$th sample is taken at time $t_i = \mbox{round}(T_{\rm uni})$.

\begin{figure}
	\centering
    \includegraphics[width=8.5cm]{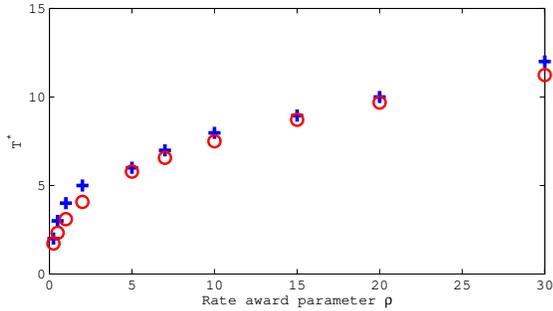}
    \caption{Illustration of Corollary~\ref{cor:stat} by simulation. The red curve is the solution of \eqref{eq:root}. The blue curve is optimal sampling rate $T^*$. Note that, their difference is always less than 1\@.}
    \label{fig:root}
  \end{figure}

First, we consider an autoregressive signal model introduced in \eqref{eq:signal-model-1}. In Theorem~\ref{thm:stat}, we show that an optimal sampling scheme for this signal is uniform. Also, Corollary~\ref{cor:stat} provides a straightforward way to compute the optimal sampling rate. Figure~\ref{fig:root} illustrates this corollary for the case of $\alpha=0.99$ (i.e., noise power is $0.02$) and for different $\rho$ values. As shown in this figure, the difference between the root of equation \eqref{eq:root} and the optimal increment $T^{*}$ is always less than 1\@. 

\begin{figure}
	\centering
    \includegraphics[width=8.5cm]{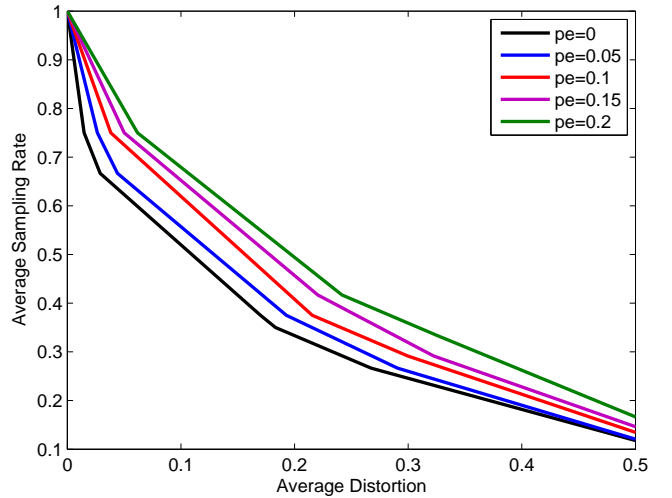}
    \caption{Analytical rate-distortion curves of Theorem \ref{thm:greedy-markovian} for a Markovian signal of equation \eqref{eq:signal-model-2} with the signal parameters $\alpha_0=0.01$ and $\alpha_1=0.99$. $P_e$ represents error probability of the estimation process. $P_e=0$ curve corresponds to the so-called genie-aided scheme.}
    \label{fig:analytical}
  \end{figure}

\begin{figure}
	\centering
    \includegraphics[width=8.5cm]{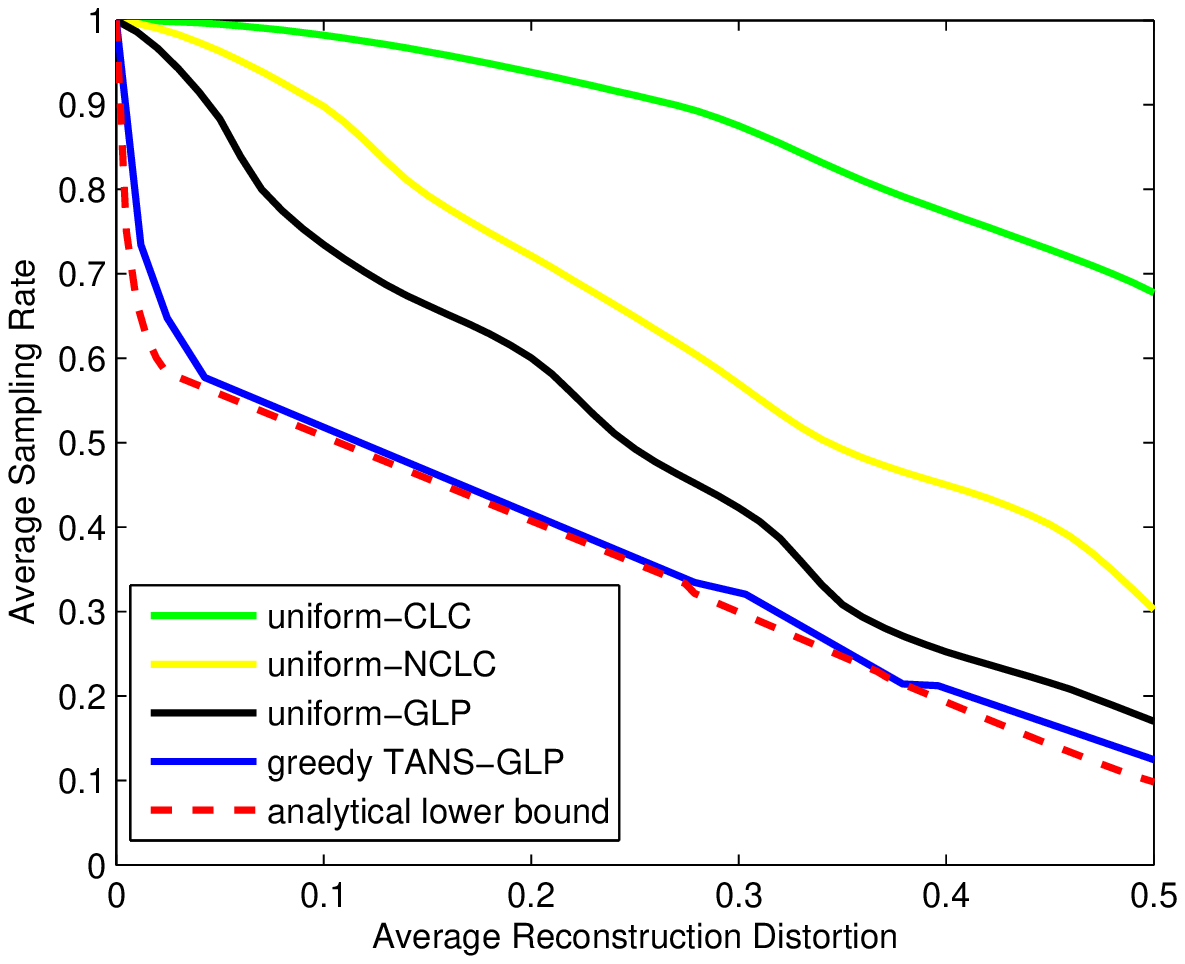}
    \caption{Average sampling rate versus average reconstruction distortion for a Markovian signal with parameters $\alpha_0=0.01$ and $\alpha_1=0.99$ for methods: (i) uniform sampling with causal line-connecting (CLC) reconstruction, (ii) uniform sampling with non-causal line-connecting (NCLC) reconstruction, (iii) uniform sampling with generalized linear prediction (GLP) filtering (iv) greedy TANS with generalized linear prediction (GLP) filtering, and (v) analytical lower bound for greedy TANS based on Theorem \ref{thm:greedy-markovian}.  }
    \label{fig:greedy1}
  \end{figure}

\begin{figure}
	\centering
    \includegraphics[width=8.5cm]{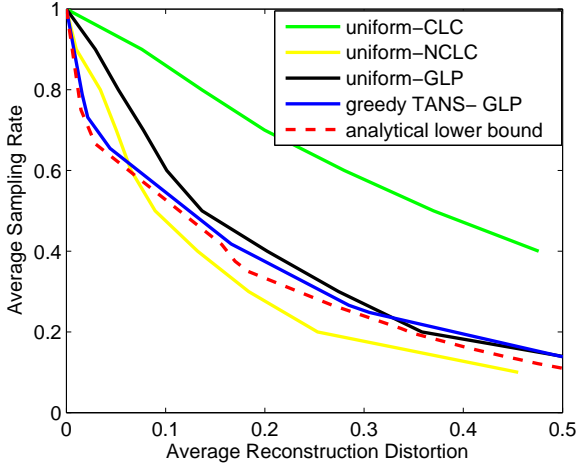}
    \caption{Average sampling rate versus average reconstruction distortion for a Markovian signal with parameters $\alpha_0=0.7$ and $\alpha_1=0.97$ for methods: (i) uniform sampling with causal line-connecting (CLC) reconstruction, (ii) uniform sampling with non-causal line-connecting (NCLC) reconstruction, (iii) uniform sampling with generalized linear prediction (GLP) filtering (iv) greedy TANS with generalized linear prediction (GLP) filtering, and (v) analytical lower bound for greedy TANS based on Theorem \ref{thm:greedy-markovian}.}
    \label{fig:greedy2}
  \end{figure}

Now, we consider Markovian signals introduced in \eqref{eq:signal-model-2}. Transition probabilities of the underlying MC are assumed to be $0.001$. We demonstrate the performance of different sampling methods on a rate-distortion plots (Figures~\ref{fig:analytical}, \ref{fig:greedy1} and~\ref{fig:greedy2}). Rate refers to average sampling rate and distortion refers to average reconstruction distortion per sample. Lower curves in these plots indicate better performance. 

Figure \ref{fig:analytical} shows analytical rate-distortion curves of Theorem~\ref{thm:greedy-markovian} for a Markovian signal of equation \eqref{eq:signal-model-2} for different estimation error probabilities. Here, noise power in state 0 of the Markov chain is 0.05 (i.e., $\alpha_0\approx 0.97$) and the noise power in state 1 is 0.5 (i.e., $\alpha_1\approx 0.7$). As illustrated in this plot, the lower the error probability, the better the performance. Note that the case where $P_e=0$ is referred as a genie-aided scheme. 

Figures \ref{fig:greedy1} and \ref{fig:greedy2} show rate-distortion curves achieved by simulations for various schemes. The signal model is Markovian \eqref{eq:signal-model-2} with parameters $\alpha_0=0.01$ and $\alpha_1=0.99$ in schemes considered in Figure \ref{fig:greedy1} and, $\alpha_0=0.7$ and $\alpha_1=0.97$ in the ones of Figure \ref{fig:greedy2}. In greedy schemes, we use a maximum likelihood estimation block with $m=10$ to estimate the state of the underlying Markov chain. For greedy TANS, we use generalized linear prediction (GLP) filter as the reconstruction method. Note that, this reconstruction is causal. For uniform sampling, we use three reconstruction methods: causal line-connecting (CLC), non-causal line-connecting (NCLC) and GLP filtering. Although, it is not fair to compare performance of greedy TANS with causal reconstruction with a uniform sampling scheme with non-causal reconstruction, in the case of Figure~\ref{fig:greedy1}, greedy TANS outperforms uniform sampling schemes including the one with a non-causal reconstruction method.
In the case of Figure~\ref{fig:greedy2}, greedy TANS outperforms uniform schemes with causal reconstructions. In the low-distortion regime (distortion less than $0.08$), it also outperforms uniform sampling with non-causal reconstruction. 

As illustrated in these figures, genie-aided greedy TANS provides an analytical lower bound for greedy TANS in the rate-distortion plot. The proposed greedy TANS performs closely to this lower bound. Also, by choosing the estimation error probability $0.05$ (estimated from simulations), an analytical upper bound for greedy TANS can be achieved as proposed in Theorem~\ref{thm:greedy-markovian}. Moreover, by comparing two greedy TANS schemes of these figures, we notice that the more different $\alpha$ values of MC states, the more gain is provided by TANS framework.  

\begin{figure}
	\centering
    \includegraphics[width=8.5cm]{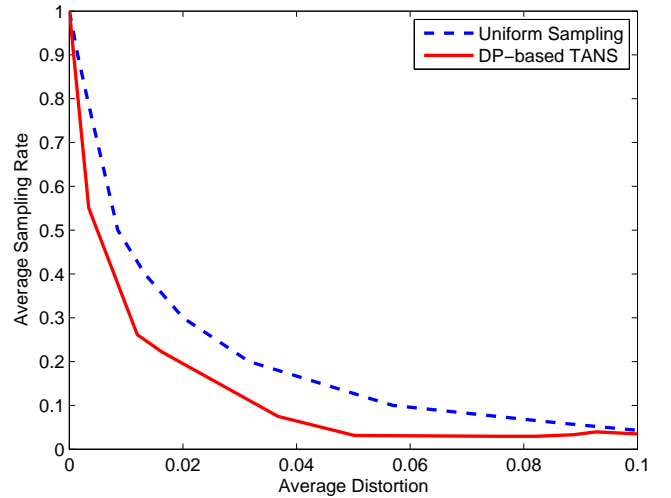}
    \caption{A rate-distortion plot of a dynamic programming-based TANS for an online source coding application explained in Section~\ref{subsec:online-source-coding}.}
    \label{fig:source-coding}
  \end{figure}

Figure~\ref{fig:source-coding} shows the performance of a dynamic programming-based TANS scheme for an online source coding application explained in Section~\ref{subsec:online-source-coding}. Here, we assume $\eps_0=0.1$ and $\eps_1=0.01$. To solve the Bellman-Ford equation \eqref{eq:bfe-source-coding}, a value-iteration method is used~\cite{dpbook}. As shown in this plot, a DP-based TANS scheme outperforms uniform sampling. 

\begin{figure}
	\centering
    \includegraphics[width=8.5cm]{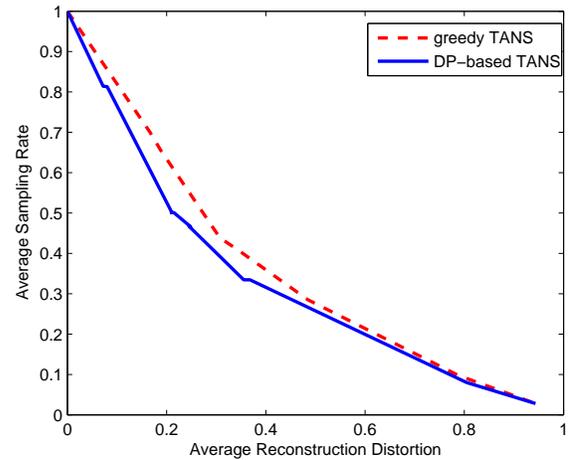}
    \caption{Comparison of a dynamic programming-based TANS with greedy TANS for a Markovian signal model.}
    \label{fig:dp-greedy}
  \end{figure}

Figure~\ref{fig:dp-greedy} illustrates performance of a TANS scheme based on approximate dynamic programming for a Markovian signal explained in Algorithm \ref{alg:adp-tans}. Here, we assume that underlying Markov chain transition probabilities are 0.1 (i.e., $p_{01}=p_{10}=0.1$). The signal parameters are assumed to be $\alpha_0=0.7$ and $\alpha_1=0.99$. As illustrated in this figure, a TANS scheme based on ADP outperforms the greedy one. 

\section{Proofs}\label{sec:proof}

\subsection{Proof of Theorem~\ref{thm:weights}}
\label{sec:proof-thm:weights}
To find a solution of the optimization problem \eqref{eq:opt-gen-linear-filter}, we use similar techniques as for the regular linear prediction filter \cite{Haykin:96}. Note that in an optimal scheme the error term should be orthogonal to all known samples:  
\begin{equation}\label{eq:ortog}
E\big[X(t_{i-k}) e^{*}(t)\big]=0 
\end{equation}
for $k=0,\ldots,m-1$, where $e^{*}(t)=X(t)-\sum_{k=0}^{m-1} w_{{\tau_k}}^{*} X(t_{i-k})$. 

By using \eqref{eq:ortog} and \eqref{eq:auto}, optimal weights $w_{\tau_i}^*$, for $0 \leq i\leq m-1$ should satisfy the following set of linear equations:
\begin{equation}\label{eq:equations}
r(-\tau_k)=\sum_{i=0}^{m-1} w_{\tau_i}^{*} r(\tau_i-\tau_k)
\end{equation}
 for $k=0,\ldots,m-1$.  By using matrix notations of \eqref{eq:matrix}, linear equations of \eqref{eq:equations} can be written as follows:
\begin{equation}
\bp=\bR \bw_{\btau}^{*}.\nonumber 
\end{equation}

\subsection{Proof of Theorem~\ref{thm:gen-linear}}
\label{sec:proof-thm:gen-linear}
Since $X(t)$ has zero mean and $X(t)=\Xh(t)+e^*(t)$, by using \eqref{eq:ortog}, we have
\begin{equation}\label{eq:sigma1}
\sigma_{e^{*}}^2(\sni,T)=\sigma_{X}^2-\sigma_{\Xh}^2
=r(0)-\sigma_{\Xh}^2,
\end{equation}
where $\Xh(t)= (\bw_{\btau}^*)^T (\bX_{\sni})$. Therefore,
\begin{eqnarray}
\sigma_{\Xh}^2&=&E[|\Xh(t)|^2] \nonumber \\
&=&(\bw_{\btau}^*)^T E\big[(\bX_{\sni}) (\bX_{\sni})^T\big]\bw_{\btau}^*\nonumber\\
&=&(\bw_{\btau}^*)^T \bR \bw_{\btau}^*\nonumber\\
&=&\bp^{T}\bw_{\btau}^{*}.
\label{eq:sigma2}
\end{eqnarray}
Equations \eqref{eq:sigma1} and \eqref{eq:sigma2} establish the theorem.

\subsection{Proof of Theorem~\ref{thm:stat}}
\label{sec:proof-thm:stat}
By using the definition of an AR(1) signal, the sampling state cost can be written as follows:
\begin{equation}
c(\sni,T)=\sum_{j=1}^{T-1} (1-\alpha ^{2j})+\frac{\rho}{T}.
\end{equation}
Hence, an optimal greedy sampling solution for this sampling state cost can be computed as follows:
\begin{equation}
\Tsgreedy=\argmin_{T} \sum_{j=1}^{T-1} (1-\alpha^{2j})+\frac{\rho}{T}.
\end{equation}

Now, we show that this optimal greedy TANS (which is a uniform sampling scheme) performs closely to an optimal sampling scheme, which may be nonuniform.
Consider a uniform sampling state $\sni$ with an inter-state sampling increment $\Tsgreedy$. The sampling state cost of this uniform sampling scheme is referred to by $\cmin$ (this sampling state cost can be achieved by having $\Tsgreedy$ as a solution of the optimization setup \eqref{eq:greedy}). First, we show that this uniform sampling scheme satisfies the BFE \eqref{eq:BFE}. Since under this sampling scheme, $S_{t_{i+1}}=\sni$, therefore, the cost-to-go function at the sampling state $\sni$ can be written as follows: 
\begin{eqnarray}
J_{f}(\sni)&=&c(\sni,T_i)+\beta c(S_{t_{i+1}},T_{i+1})+\cdots\nonumber\\
&=& c(\sni,\Tsgreedy)\big(1+\beta+\beta^2+\cdots\big)\nonumber\\
&=& \frac{\cmin}{1-\beta}. 
\end{eqnarray}

Now, consider the right hand side (RHS) of the BFE \eqref{eq:BFE}:
\begin{eqnarray}
\mbox{RHS of BFE}&=&\min_{T_i} \big(c(\sni,T_i)+\beta E[J_{f}(S_{t_{i+1}})]\big)\nonumber\\
&=&\cmin+\beta \frac{\cmin}{1-\beta}\nonumber\\
&=&\frac{\cmin}{1-\beta}\nonumber\\
&=&J_{f}(\sni).
\end{eqnarray}
Therefore, the above uniform sampling scheme satisfies the BFE when $\sni$ (i.e., the initialization state) happens to be a uniform sampling state with an inter-state sampling step size $\Tsgreedy$. If the process starts from another sampling state, we assume that there is always a way to reach to this uniform sampling state (e.g., take $m$ uniform samples with a sampling increment $\Tsgreedy$). Moreover, the mapping functions between policies and costs are continuous. Hence, a small difference in costs due to initialization effects has a small effect in sampling policies. Therefore, for an AR(1) signal model, the optimal greedy TANS (which is uniform) performs closely to an optimal sampling scheme (which may be nonuniform).

\subsection{Proof of Corollary~\ref{cor:stat}}
\label{sec:proof-cor:stat}
Define $h(T)=(1-\alpha^{2T})-\Frac{\rho}{(T(T+1))}$ for $T \in [1,\infty)$. Note that $h(T)$ is a continuous function over this interval. Since $h(1)=(1-\alpha^2)-\Frac{\rho}{2}<0$ and $\lim_{T\to\infty}h(T)=1>0$, by using the mean value theorem, $h$ has a root $\Troot$ in $[1,\infty)$. Since $\Frac{d h(T)}{d T}>0$, this root is unique, completing the proof of part (i).

For an AR(1) signal, we have:
\begin{equation}
h(T)=c(\sni,T+1)-c(\sni,T)=(1-\alpha^{2T})-\frac{\rho}{T(T+1)}.\nonumber
\end{equation}
Since $T^*=\argmin_{T} c(\sni,T)$ and also $T^*$ is an integer, either $T^*=\lfloor \Troot\rfloor$ or $T^*=\lfloor\Troot\rfloor+1$, completing the proof of part (ii).

\subsection{Proof of Theorem~\ref{thm:greedy-markovian}}
\label{sec:proof-thm:greedy-markovian}
Under conditions of \eqref{eq:conditions-markovian}, \eqref{eq:cost-markov-apx} shows the sampling state cost given $\htsni=0$. By using the sampling scheme proposed in Algorithm~\ref{def:greedy-markovian}, we have: 
\begin{equation}
T_i=\argmin_{T} c(\sni,T|\htsni=0).
\end{equation}
Since $\maxervar\geq 1$, by having $P_e^{\rm low}\leq P(\sni)\leq P_e^{\rm up}$, given $\htsni=0$, we have: $T_0^{\rm low}\leq T_i\leq T_0^{\rm up}$. Similarly, one can show that, given $\htsni=1$, $T_1^{\rm low}\leq T_i\leq T_1^{\rm up}$. Since the underlying Markov chain is symmetric (i.e., $p_{01}=p_{10}$), in steady state, $\htsni=0$ approximately half of the time and $\htsni=1$ approximately half of the time. Hence, the number of samples taken in that state by using the proposed sampling scheme of Algorithm~\ref{def:greedy-markovian} is bounded between
$$
\frac{\Ttot}{2T_0^{\rm low}}+\frac{\Ttot}{2T_1^{\rm low}}
\qquad \mbox{and} \qquad
\frac{\Ttot}{2T_0^{\rm up}}+\frac{\Ttot}{2T_1^{\rm up}}.
$$
This demonstrates the sampling rate bounds. By using bounds on $T_i$ and $P_e(\sni)$ in \eqref{eq:cost-markov-apx}, deriving bounds on the expected reconstruction distortion is straightforward.  


\section{Conclusion}\label{sec:conc}

In this paper, we introduced a new framework for an adaptive nonuniform sampling scheme called \emph{time-stampless adaptive nonuniform sampling} (TANS). The key idea of this framework is that, \textit{time increments between samples are computed by using a function of previously taken samples}. Therefore, keeping sampling times (time stamps), except initialization times, is not necessary. We introduced two methods to design sampling functions for discrete-time stochastic signals:
a greedy method, and a method based on dynamic programming.
We analyzed the performances of these schemes by computing (or bounding) their trade-offs between sampling rate and expected reconstruction distortion for autoregressive and Markovian signals. We showed that, by being time-stampless and opportunistically adapting to local signal characteristics, TANS can provide significant rate-distortion gains, which can be translated to improved power efficiency in some applications.

\section{Acknowledgment}

Authors would like to thank Muriel Rambeloarison for helping in some of the simulations.


\end{document}